\documentclass[12pt]{article}

\usepackage{amsfonts}
\usepackage{amsmath}

\usepackage{eucal}
\usepackage{amssymb}

\newcommand{\eq}{\begin{equation}}
\newcommand{\feq}{\end{equation}}
\newcommand{\eqn}{\begin{eqnarray}}
\newcommand{\feqn}{\end{eqnarray}}
\newcommand{\meq}{\begin{multline}}
\newcommand{\fmeq}{\end{multline}}
\newcommand{\arr}{\begin{eqnarray*}}
\newcommand{\farr}{\end{eqnarray*}}

\newcommand{\DD}{{\cal D}}

\csname @addtoreset\endcsname{equation}{section}

\def\rmd{{\rm d}}

\begin{document}
\begin{titlepage}
\begin{flushright}
IFUM-720-FT \\

\end{flushright}
\vspace{.3cm}
\begin{center}
\renewcommand{\thefootnote}{\fnsymbol{footnote}}
{\Large \bf BPS equations in ${\cal N}=2$, $D=5$ supergravity
with hypermultiplets}
\vskip 15mm
{\large \bf {Sergio~L.~Cacciatori$^{1,3}$\footnote{cacciatori@mi.infn.it}
Alessio~Celi$^{2,3}$\footnote{alessio.celi@mi.infn.it} and
Daniela~Zanon$^{2,3}$\footnote{daniela.zanon@mi.infn.it}}}\\
\renewcommand{\thefootnote}{\arabic{footnote}}
\setcounter{footnote}{0}
\vskip 10mm
{\small
$^1$ Dipartimento di Matematica dell'Universit\`a di Milano,\\
Via Saldini 50, I-20133 Milano, Italy\\

\vspace*{0.5cm}

$^2$ Dipartimento di Fisica dell'Universit\`a di Milano,\\
Via Celoria 16, I-20133 Milano, Italy\\

\vspace*{0.5cm}

$^3$ INFN, Sezione di Milano,\\
Via Celoria 16,
I-20133 Milano, Italy\\
}
\end{center}
\vspace{2cm}
\begin{center}
{\bf Abstract}
\end{center}
{\small } With the general aim to classify BPS solutions in ${\cal N}=2$, $D=5$ supergravities interacting 
with an arbitrary number of vector, tensor and hypermultiplets, here we begin  considering the most 
general electrostatic, spherical-symmetric BPS solutions
in the presence of hypermultiplet couplings.
We discuss the properties of the BPS equations and the restrictions imposed by their integrability conditions. 
\end{titlepage}

\section{Introduction}
In recent years five-dimensional ${\cal N}=2$ gauged supergravity theories have received considerable 
attention primarily for their relevance  to the AdS/CFT correspondence \cite{ads}. In particular much 
interest has been directed toward the study of domain-wall supergravity solutions \cite{flat-domainwall}, 
\cite{curved-domainwall} as duals of renormalization group (RG) flows in the corresponding  field theory 
\cite{RGF}. Also a  strong motivation in this direction derives from phenomenological requests in 
brane-world scenarios obtained via M-theory compactifications and/or domain-wall type models \cite{phenom}.

Finding supersymmetric solutions of ${\cal N}=2$, $D=5$ supergravities is never an easy exercise \cite{Gauntlett:2002nw}; it 
becomes a quite difficult task if one considers general couplings to matter and general gaugings. Partial 
results have been obtained so far, i.e. for cases where only special vector or hypermultiplet gaugings 
have been considered \cite{partialsolut}, \cite{gutperle:2001}. Here we start a systematic program with the general aim to classify BPS solutions 
with vector, tensor and hypermultiplet couplings. 
The introduction of the hypermultiplets is crucial for widening the variety of solutions as compared to the case where only vector multiplets are present \cite{ceresole:2001}, \cite{vectorsol}. In particular, 
aside for the special example analyzed in \cite{gutperle:2001},
the existence of BPS  black-hole  solutions has not been investigated systematically. Of course,
 black-hole solutions would be especially relevant since, via the AdS/CFT correspondence, they could describe the RG flows between field theories in different dimensions \cite{maldacena:2000}.   

In this paper we restrict ourselves to the case of 
hypermultiplet couplings, with generic gauging, and a static $SO(4)$ symmetric ansatz for the metric. 
In this setting we study the 
integrability conditions that follow from the BPS equations and find a set of equations
for the functions in the ansatz. The quaternionic geometries give equations for the scalars  which 
are a generalization of the ones found in \cite{ceresole:2001}.
Then we analyze all these equations and check directly that they 
satisfy the equations of motion.  

Our paper is organized as follows: in the next section we introduce the model and the basic ingredients. 
For the construction of the most general $\cal{N}$=2 gauged supergravities and for technical details 
we refer the reader to  the existing literature \cite{ceresole:2000} and to references therein. We describe 
the form of the solutions we are looking for: obviously this choice determines 
the {\em physics} contained in the solution. 
Then we focus on the derivation of 
the BPS equations and study their 
integrability conditions.
In section \ref{blackhole}
we find the set of independent first order differential equations that are equivalent to the BPS 
conditions. In section \ref{equamotion} we show that the family of solutions we have found satisfy 
the equations of motion. In section \ref{solution} we discuss the properties of the BPS solutions 
in the special case of the universal hypermultiplet \cite{witten:1983} and find an explicit result 
for a simple choice of the gauging.
 We conclude with some final remarks.
Our notations and conventions are summarized in an Appendix. 

\section{\large The model and its BPS equations}\label{model}
We consider $\cal{N}$=2 gauged supergravities in five dimensions interacting with an arbitrary number of 
hypermultiplets. (In this paper we do not study vector and tensor multiplet couplings.) 
The field content of 
the theory is the following
\begin{itemize}
\item the supergravity multiplet \eqn \{ e^a_\mu \ , \psi^{i}_\mu \ , A_\mu \} \feqn containing the 
{\em graviton} $e^a_\mu$, two {\em gravitini} $\psi^{i}_\mu$ and the {\em graviphoton} $A_\mu$,
which is the only (abelian) gauge field present in the theory; \item $n_H$ hypermultiplets 
\eqn \{ \zeta^A \ , q^X \} \feqn containing the {\em hyperini} $\zeta^A $ with ${\scriptstyle A}=1,2,\dots,2 n_H$, and the {\em scalars} $q^X$ with ${\scriptstyle X}=1,2,\dots,4 n_H$ which define a quaternionic K\"ahler manifold\footnote{We collect the basic geometrical notions in the Appendix.}  with 
metric $g_{XY}$.
\end{itemize}

 \noindent The bosonic sector of the theory is described by the Lagrangian density presented in
\cite{ceresole:2000}
\eqn \label{lagr}        
&&{\cal {L}}_{BOS} =\ - \ \frac 12 e \left[ R+\frac 12 F_{\mu \nu} F^{\mu \nu} 
                  +g_{XY} D_\mu q^X D^\mu q^Y \right] \cr
&&~~~~~~~~~~~~
+ \frac 1{6\sqrt 6} \epsilon^{\mu\nu\rho\sigma\tau} F_{\mu\nu}           F_{\rho\sigma}A_{\tau} 
- e {\cal V} (q)  
\feqn
with 
$$
D_\mu q^X=\partial_{\mu} q^X + g A_\mu K^X(q)
$$
where $K^X(q)$ is a Killing vector on the quaternionic manifold and ${\cal V} (q)$ is the scalar potential as given in Appendix.

We look for electrostatic spherical solutions that preserve half of the ${\cal N}=2$ supersymmetries. 
To this end we make the following ansatz for the supergravity fields: we choose a metric which is $SO(4)$ symmetric with all the other fields that only depend on the holographic space-time coordinate $r$. Moreover we fix the gauge for the graviphoton keeping only the $A_t$ component different from zero. 
 
Introducing  spherical coordinates $(t,r,\theta ,\phi, \psi)$ we write 
\cite{gutperle:2001}
 \eqn \label{metricansatz} ds^2 =-e^{2v} dt^2 +e^{2w}  dr^2 + r^2 
(d\theta^2 +\sin^2 \theta d\phi^2 +\cos^2 \theta d\psi^2 ) \feqn 
where the functions $v$ and $w$ depend on $r$ only. The variations of the fermionic fields under supersymmetry 
transformations give rise to the following BPS equations: for the gravitini we have 
\cite{vanproeyen:2001} 
\eqn 0 \ = \ \delta_\epsilon \psi_{\mu i} &=& \partial_\mu \epsilon_i +\frac 
14 \omega^{ab}_\mu \gamma_{ab} \epsilon_i -\partial_\mu q^X p_{Xi}^{\quad j} \epsilon_j -  g 
A_{\mu} P_{i}^{\ j} \epsilon_j \cr &+&\frac i{4 \sqrt 6} (\gamma_{\mu \nu \rho} -4g_{\mu \nu} 
\gamma_\rho )F^{\nu \rho } \epsilon_i - \frac i{\sqrt 6} g  P_{i}^{\ j} \gamma_\mu \epsilon_j  
\label{BPSgrav} \feqn 
We note that in ref.\cite{gutperle:2001}  the corresponding equation contains
an additional term. This extra term, which arises due to a incorrect interpretation of the covariant 
derivative acting on the spinor
$\epsilon$ as given in \cite{ceresole:2000},  should not be present. 

For the hyperini the equations $ \delta_\epsilon \zeta^A=0$ lead to
 \eqn \left[\frac i2 e^{-w} f^{Ai}_X 
q^{\prime X} \gamma_1 - i \frac g2 f^{Ai}_X K^X e^{-v} A_t \gamma_0 \right] \epsilon_i &=&
 \frac {\sqrt 6}4 g K^X f^{Ai}_X \epsilon_i
\label{BPShyper}  \feqn where we have set $q^{\prime X} =\partial_r q^X$

Without loss of generality it is convenient to parametrize the graviphoton as follows \footnote{Note that in \cite{gutperle:2001}  $a$ was chosen to be a constant.}
\eqn \label{elettrico} A_t =\sqrt {\frac 32} a(r)e^v \feqn 
This allows to write explicitly  the BPS equations for the gravitini 
\eqn & & \left\{
\partial_t \delta_l^{\ k} +\frac 12 v^\prime e^{v-w} \delta_l^{\
k} \gamma_0 \gamma_1 +\frac 1{\sqrt 6} g e^v P^r (\sigma_r )_l^{\
k} \gamma_0 +\frac i2 e^{v-w} (v^\prime a+ a^\prime)  \delta_l^{\ k}
\gamma_1 \right. \cr
& &~~~~~~ \left. -i\sqrt{\frac 32}g a e^v  P^r  (\sigma_r )_l^{\ k} \right\}
\epsilon_k =0
\label{tBPSgrav} \\
&&~~~\cr
&&~~~\cr
& & \left[ \partial_r \delta_i^{\ j} - iq^{\prime X} p_X^r (\sigma_r )_i^{\ j}
+\frac 12 (v^\prime a + a^\prime) (i\gamma_0 )\delta_i^{\ j}
+\frac {g} {\sqrt 6} e^w P^r \gamma_1 (\sigma_r )_i^{\ j} \right] \epsilon_j =0
\label{rBPSgrav} \\
&&~~~\cr
&&~~~\cr
& & \left[
\delta_i^{\ j} \partial_\theta -\frac 12 e^{-w}
\gamma_1 \gamma_2 \delta_i^{\ j} +i\frac {re^{-w}}4 (v^\prime a + a^\prime)
\gamma_{012}
\delta_i^{\ j} 
 +\frac {g r}{\sqrt 6} P^s (\sigma_s)_i^{\ j} \gamma_2
\right] \epsilon_j =0 \label{teta}  \\
&&~~~\cr
&&~~~\cr
& & \left\{
\delta_i^{\ j} \left[ \partial_\phi -\frac 12 e^{-w}
\sin \theta \gamma_1 \gamma_3 -\frac 12 \cos \theta \gamma_2 \gamma_3
+i\frac {re^{-w}}4 (v^\prime a + a^\prime) \sin \theta \gamma_{013} \right] \right.
\cr
& & ~~~~~~\left. +\frac {g r \sin \theta}{\sqrt 6} P^s (\sigma_s)_i^{\ j}
\gamma_3
\right\} \epsilon_j =0 \label{fi}    \\
&&~~~\cr
&&~~~\cr
& & \left\{ \delta_i^{\ j} \left[ \partial_\psi -\frac 12 e^{-w} \cos \theta \gamma_1 
\gamma_4 +\frac 12 \sin \theta \gamma_2 \gamma_4 +i\frac {re^{-w}}4 (v^\prime a + a^\prime) \cos \theta 
\gamma_{014} \right] \right.  \cr 
& & ~~~~~~\left. +\frac {g r \cos \theta}{\sqrt 6} P^s (\sigma_s)_i^{\ j} 
\gamma_4 \right\} \epsilon_j =0 \label{psi} \feqn 

At this point, using (\ref{fquadrato}) and the $SU(2)$ projection as in (\ref{pro}), we can rewrite the 
algebraic relations in (\ref{BPShyper})  as
\eqn  \label{1BPShyper}
\left( i e^{-w} q^{\prime X} \gamma_1 -i g\sqrt{\frac 32} a K^X
\gamma_0 - \sqrt{\frac 32} g K^X \right) \left(g_{ZX} \delta_j^{\ i} +
2 i R_{ZX}^s {(\sigma_s)}_j^{\ i}\right) \epsilon_i =0
\feqn
We will make use of the above expression in the following.

\subsection{\large Integrability conditions}

Now we want to discuss the integrability of the gravitini equations in order to insure the existence of a 
Killing spinor (i.e of residual supersymmetry). The standard procedure is to impose the vanishing of the 
various commutators. In this way one obtains equations that combined with the hyperini ones determine the 
unknown functions in the ansatz and impose restrictions on the geometry (gauging). We find it useful to 
adopt the following notation: given the vector $P^s$  $s=1,2,3$ we introduce the phase $\vec Q$ 
so that $\vec Q \cdot \vec Q =1$ and use the decompositions of the vector into its norm and phase
\eqn
\vec P =\sqrt {\frac 32} W \vec Q
\ .
\label{1scomposizione}
\feqn
We find four independent\footnote{Symmetry arguments show that the angular 
equations lead to only one independent condition.} integrability conditions that we list below:\\
from the commutators between the BPS equations (\ref{teta}), (\ref{fi}) and (\ref{psi}) we obtain
\eqn
& &
\left\{ i\gamma_0 \left[ 1-e^{-2w}
 -\frac {r^2 e^{-2w}}4 (v^\prime a + a^\prime)^2  +{g}^2
 r^2 W^2 \right] \delta_i^{\ j} \right. \cr
& & \Bigl.
+\bigl[ re^{-2w}\left(v^\prime a + a^\prime\right) \bigr]
\delta_i^{\ j} +\gamma_1 \bigl[ g e^{-w} r^2  \left(v^\prime a + a^\prime\right)
W Q^s \bigr] (\sigma_s)_i^{\ j} \Bigr\} \epsilon_j =0\cr
&&~~~~~
\label{tetafi}
\feqn
The commutators between (\ref{tBPSgrav}) and
the angular components give the conditions
\eqn
& & \Bigl\{
\gamma_0 \left[ v^\prime e^{-w} -{g}^2 e^{w} r
W^2 \right] \delta_i^{\ j} -i g r (v^\prime a+ a^\prime) WQ^s
(\sigma_s)_i^{\ j} \gamma_1 
 -i(v^\prime a+ a^\prime) e^{-w} \delta_i^{\ j}
 \Bigr\} \epsilon_j =0 \cr
&&~~~~~
\label{tteta}
\feqn
The commutators between (\ref{rBPSgrav}) and the angular 
components give
\eqn
& & \left\{ -\frac i2 g (v^\prime a+ a^\prime) 
  rW Q^s (\sigma_s)_i^{\ j} \gamma_0 +\left[ \frac 12
-w^\prime e^{-w}  -\frac {{g}^2}2 e^{w} rW^2
\right] \gamma_1 \delta_i^{\ j} \right. \cr
& & \left. +\frac {g}2 r {q^\prime}^X D_X \left( WQ^s \right) (\sigma_s)_i^{\ j} 
+\frac i4 \partial_r \left[ re^{-w} (v^\prime a + a^\prime)
\right] \gamma_0 \gamma_1 \delta_i^{\ j}  \right\} \epsilon_j =0
\label{rteta}
\feqn
Finally the commutator between (\ref{tBPSgrav}) and
(\ref{rBPSgrav}) gives
\eqn
& & \left\{  
ig{q^\prime}^X D_X \left(
\sqrt{\frac 32} e^v a WQ^s \right) (\sigma_s)_i^{\ j}
-\gamma_0 \frac {g}2 e^v {q^\prime}^X D_X(WQ^s ) (\sigma_s)_i^{\ j}
\right. \cr
& & +\gamma_1 \left[ i \frac {v^\prime}2  e^{v-w}
(v^\prime a + a^\prime) \delta_i^{\ j} 
-\frac i2 \partial_r \left(
e^{v-w} (v^\prime a + a^\prime) \right) \delta_i^{\ j} \right] \cr
& & \left. +\gamma_0 \gamma_1 \delta_i^{\ j} \left[ \frac 12 {g}^2
e^{v+w} W^2 +e^{v-w} \frac 12 (v^\prime a+ a^\prime)^2 -\frac 12
\partial_r (v^\prime e^{v-w}) \right] \right\} \epsilon_j =0
\label{tierre}
\feqn

We begin with the study of the equation in (\ref{tetafi}). Unless all
coefficients vanish it can be written as 
\eqn \label{ansatz}
(if^0 \gamma_0 \delta_l^{\ k} +f^r (\sigma_r )_l^{\
k} \gamma_1 )\epsilon_k =\epsilon_l
\feqn
where
\eqn
& & f^r=-g e^w rWQ^r \label{fr} \\
& & \cr
& & f^0=-\frac {1-e^{-2w} -\frac
{r^2 e^{-2w}}4 (v^\prime a + a^\prime)^2 +{g}^2 r^2 W^2}{r
e^{-2w} (v^\prime a + a^\prime) } \label{fo}
\feqn
The Killing equation in (\ref{ansatz}), viewed as a projector equation, leads to the consistency requirement 
\eqn
(f^0 )^2 +\sum_r (f^r )^2 =1 \label{normal}
\feqn
Now we compare the above results with the content from the hyperini equation in (\ref{1BPShyper}). 
Starting from (\ref{1BPShyper}),  multiplying by the projector
$K_{\tilde Z} \delta_i^{\ j} -2iR^s_{\tilde Z X} K^X (\sigma_s)_i^{\ j}$
and symmetrizing in ${\scriptstyle Z, \  \tilde Z}$, we obtain
\eqn
& & \left[ \Bigl(K_Z K_{\tilde Z} +4K^X K^Y R^r_{ZX} R^s_{\tilde Z Y} \delta_{rs} \Bigr)
\Bigl[ a \gamma_0 \delta_i^{\ j} - i \delta_i^{\ j}\Bigr] 
-\left( \sqrt {\frac 23} \frac {e^{-w}}{g} q^{\prime
X} g_{X(Z} K_{\tilde Z )} \delta_i^{\ j} \right. \right. \cr
& & + 2 \sqrt{\frac 23} i
\frac {e^{-w}}{g} q^{\prime X} R^r_{(Z|X|} K_{\tilde Z )} (\sigma_r)_i^{\ j}
-2 \sqrt{\frac 23} i \frac {e^{-w}}{g} q^{\prime X} g_{X(Z}
R^r_{\tilde Z )Y} K^Y (\sigma_r)_i^{\ j} \cr
& & \left. \left. +4 \sqrt {\frac 23} \frac {e^{-w}}{g} q^{\prime X}
R^r_{(Z |Y|} R^s_{\tilde Z )X} K^Y \Bigl(\delta_{rs} \delta_i^{\ j}
+i \epsilon_{rst} (\sigma_r)_i^{\ j} \Bigr) \right) \gamma_1 
 \right] \epsilon_j =0 \label{w}
\feqn
The above equation is compatible with (\ref{ansatz}) only if the conditions
\eqn
q^{\prime X} (g_{X(Z} K_{\tilde Z )} -4R^r_{X(Z} R^s_{\tilde Z )Y} K^Y
\delta_{rs})=0 \label{conditions}
\feqn
are satisfied. 
If this is the case then (\ref{w}) becomes
\eqn
\left[ -ia \gamma_0 \delta_i^{\ j} - 2 \sqrt{\frac 23} e^{-w} q^{\prime X}
\frac {U^r_{XZ\tilde Z}}{g( K_Z K_{\tilde Z} +4K^T K^Y R^r_{ZT} R^s_{\tilde Z Y}
\delta_{rs}) } \gamma_1 (\sigma^r)_i^{\ j} \right] \epsilon_j =\epsilon_j
\label{confronto}
\feqn
with
\eqn
U^r_{XZ\tilde Z} \equiv R^r_{(Z|X|} K_{\tilde Z )} -g_{X(Z} R^r_{\tilde Z )Y} K^Y
+2 R^t_{X(Z} R^s_{\tilde Z )Y} K^Y \epsilon_{tsr}
\feqn
The second term in (\ref{confronto}) must be independent of $Z$ and $\tilde Z$, so that one obtains   
\eqn
q^{\prime X}\frac {U^r_{XZ\tilde Z}}{K_Z K_{\tilde Z} +4K^T K^Y R^t_{ZT}
R^s_{\tilde Z Y}\delta_{ts}} =\ -\ \frac{q^{\prime X} D_X P^r}{|K|^2}\ = \ 
-\sqrt{\frac 32} \frac {g e^w}2 f^r
\label{vincolo}
\feqn
The above equations have important consequences  for the geometry of the moduli space; we will discuss them in the next section.

At this point from the angular integrability condition and the hyperini supersymmetry variation we have
\eqn
\sqrt {\frac 32}~ g^2 e^{2w} rWQ^r
&=& -2\frac{q^{\prime X} D_X P^r}{|K|^2} \label{210} \\
f^0 &=& -a    \label{semplice}  
\feqn
The above result and the relation in (\ref{normal}) fix $f^r$ to be
\eqn
f^r =\pm \sqrt {1-a^2} Q^r  \label{parallela}
\feqn
In addition from the vector relation\footnote{$Q^rQ_r=1$ implies $(D_XQ^r)Q_r=0$}  (\ref{210})
we have
\begin{gather}
q^{\prime X} D_X Q^r =0 \label{costante} \\
ge^w |K|^2 \sqrt {1-a^2} =\pm 2W^\prime  \label{corol}
\end{gather}
If we use (\ref{semplice}) and (\ref{parallela}) in (\ref{fo}) and
(\ref{fr}) then we find
\begin{gather}
\sqrt {1-a^2} =\mp g e^w rW  \label{1fr} \\
a= \frac {1-e^{-2w} -\frac
{r^2 e^{-2w}}4 (v^\prime a + a^\prime)^2 +{g}^2  r^2 W^2}{r 
e^{-2w} (v^\prime a + a^\prime)} \label{1fo}
\end{gather} 
Finally inserting (\ref {1fr}) into (\ref{1fo}) we obtain
\eqn
1=a^2 e^{-2w} \left[ 1 +\frac {r}2 
\left( v^\prime + \frac {a^\prime} a \right) \right]^2  \label{511}
\feqn

We postpone the discussion of the other consistency conditions and
analyze next the implications of what we have just found for the geometric structure of the moduli space. 

\subsection{Geometric restrictions}

Now we want to consider the equations in (\ref{conditions}) and (\ref{vincolo}) and show that they determine the space-time $r$-dependence of the scalars, i.e. of all the quantities that enter in the description of the quaternionic geometry like the prepotential $P^r$ and the Killing vectors.
In order to elucidate their meaning in a transparent manner it is convenient to proceed as follows. First a double contraction of (\ref{conditions}) with the Killing vector leads to ${q^\prime}^X K_X =0$. Then using this result and contracting (\ref{conditions}) with $K^Z$ one obtains
\eqn
{q^\prime}^Z =\pm 3ge^w \sqrt {1-a^2} \partial^Z W \label{erreeffe}
\feqn
The above equation is quite important: it describes the path in the moduli space associated 
to the BPS solution. It shows explicitly
 that, if we exclude the case $a^2=1$, the condition to have a fixed point is $\partial_Z W=0$ which corresponds to a local minimum of the potential as observed in \cite{ceresole:2001}. 

From (\ref{erreeffe})
using $q^{\prime Z} W_Z =W^\prime$ we obtain
\eqn
|q^\prime |^2 =\pm 3ge^w\sqrt {1-a^2} W^\prime  \label{1erreeffe}
\feqn
The contraction of (\ref{conditions}) with $q^{\prime Z}$ gives
\eqn \label{condq}
|q^\prime |^2 K_Z =2\sqrt 6 \delta_{rs} q^{\prime X} R^r_{XZ} Q^s W^\prime
\feqn
Acting now with $K^Z$ one obtains
\eqn
|K|^2 |q^\prime |^2 =6W^{\prime 2} \label{1condq}
\feqn
(\ref{1condq}) together with (\ref{1erreeffe}) gives again (cfr
\ref{corol})
\eqn
|q^\prime |^2 = \frac 32 |K|^2 g^2 e^{2w} (1-a^2 )   \label{2corol}
\feqn
which is in agreement with the fact that the Killing vector $K^X$ has to be null at the fixed point.
Using (\ref{2corol}) and (\ref{erreeffe}) in (\ref{condq})
one easily obtains
\eqn
K^Z =2\sqrt 6 \delta_{rs} Q^r R^{sXZ} \partial_X W  \label{3corol}
\feqn
Now we consider  (\ref{vincolo}) rewritten as
\eqn
|K|^2 q^{\prime X} U^r_{XZ\tilde Z} =
-\left[ K_Z K_{\tilde Z} +4K^T K^Y R^t_{ZT}
R^s_{\tilde Z Y}\delta_{ts} \right]
q^{\prime X} D_X P^r
\label{1vincolo}
\feqn
Contracting with $K^{\tilde Z}$ and using $K^X D_X P^s =0$ (as follows from
the definition (\ref{prepot})) we obtain
\eqn
K^2 q^{\prime X} R^s_{ZX} =-\sqrt {\frac 32} W^\prime Q^s K_Z -3WW^\prime
Q^t D_Z Q^r {\epsilon_{tr}}^s \label{importante}
\feqn
Moreover contracting (\ref{1vincolo}) with $q^{\prime \tilde Z}$ and using
(\ref{1condq})
we find
\eqn
|q^\prime |^2 \partial^Z W ={q^\prime}^Z W^\prime
\feqn
(which can be obtained also from (\ref{2corol}) and
(\ref{3corol})) and
\eqn
|q^\prime |^2 W D_Z Q^r +2R^t_{ZX} W^\prime q^{\prime X} Q^s
{\epsilon_{ts}}^r =0
\feqn
which after use of (\ref{importante}) gives 
(\ref{1condq}). \\
Note that (\ref{prepot}) gives
\eqn
K^Z =-\frac 43 R^{rZX} D_X P^r
\feqn
so that (\ref{3corol}) can be written as
\eqn \label{I3corol}
K_Z =\sqrt 6 W R^{rX}_{\quad Z} D_X Q_r
\feqn
Finally the contraction of (\ref{importante}) with $Q^r \epsilon_{srt}$ gives
\eqn
3WW^\prime D_Z Q_t =|K|^2 Q^r q^{\prime X} R^s_{ZX} \epsilon_{srt}
\label{final}
\feqn
Note that from (\ref{erreeffe}) and (\ref{1erreeffe}) we also have
\eqn
|\partial W|^2 =\frac {|K|^2}6  \label{ultima}
\feqn
Let us collect the main results:
\eqn
& & {q^\prime}^Z =\pm 3ge^w \sqrt {1-a^2} \partial^Z W \label{restrizione} \\
& & |K|^2 |q^\prime |^2 =6W^{\prime 2} \label{1restrizione} \\
& & |q^\prime |^2 K_Z =2\sqrt 6 \delta_{rs} q^{\prime X} R^r_{XZ} Q^s
W^\prime  \label{2restrizione} \\
& & K^Z =2\sqrt 6 \delta_{rs} Q^r R^{sXZ} \partial_X W \label{3restrizione} \\
& & K^2 q^{\prime X} R^s_{ZX} =-\sqrt {\frac 32} W^\prime Q^s K_Z -3WW^\prime
Q^t D_Z Q^r \epsilon_{tr}^{~ s}  \label{4restrizione}
\feqn
From the above relations it follows
\eqn
& & |q^\prime |^2 = \frac 32 |K|^2 g^2 e^{2w} (1-a^2 ) \label{5restrizione}\\
& & K_Z =\sqrt 6 W R^{rX}_{\quad Z} D_X Q_r \label{6restrizione} \\
& & |\partial W|^2 =\frac {|K|^2}6 \label{7restrizione} \\
& & 3WW^\prime D_Z Q_t =|K|^2 Q^r q^{\prime X} R^s_{ZX} \epsilon_{srt} \\
& & K_X q^{\prime X} =0 \label{ok}
\label{8restrizione}
\feqn
Now it is straightforward to verify that these conditions solve  (\ref{conditions}), (\ref{vincolo}) 
and (\ref{1BPShyper}) identically.
Finally we note that (\ref{8restrizione}) gives
\eqn
q^{\prime X} D_X Q^r =0 \label{1costante}
\feqn
It is interesting  and not at all obvious that (\ref{erreeffe}) and the other relations we 
have found in this section look like a simple generalization of those obtained for flat domain 
wall solutions (where the gauge fields are zero). This is suggestive of an underlying general 
structure, independent of the form of the space-time solution.

\subsection{Further restrictions}

The equations obtained in the previous subsection are quite general. 
Now we have to consider the other integrability conditions together with (\ref{ansatz}).
We start from (\ref{tteta}): it is easy to show that either all the coefficients vanish or it must be equivalent to (\ref{ansatz}). The first case reduces to the case in which all the
coefficients of (\ref{ansatz}) vanish.
The second case is verified when the following conditions are true:
\eqn
& & f^0 =-\frac {v^\prime  
-{g}^2 e^{2w} rW^2}{(v^\prime a + a^\prime) } \label{questa} \\
& & \cr
& & f^r=-g e^w rWQ^r \ .
\label{verificare}
\feqn
Other consequences of (\ref{tteta}) are the following: \\
from (\ref{questa}) and (\ref{fo}) we find
\eqn
1+2{g}^2 r^2 W^2 +\frac {r^2 e^{-2w}}4 \left[{v^\prime}^2-(v^\prime a + a^\prime)^2 \right]
={\left(1+\frac {r}2 v^\prime \right)}^2 e^{-2w} \label{serve?}
\feqn
while inserting (\ref{1fr}) into (\ref{questa}) we obtain
\eqn
a= \frac {rv^\prime -1+a^2 }{r(v^\prime a + a^\prime)} \label{a}
\feqn

Now we consider the integrability condition (\ref{rteta}): by means of (\ref{ansatz}) 
we obtain the equations\footnote{These two equations give again the condition (\ref{1costante}).}
\eqn
& & g r (v^\prime a + a^\prime) W +g a rW^\prime \mp \frac 12 
 \sqrt {1-a^2} \partial_r [re^{-w}(v^\prime a + a^\prime)]=0 \ , \label{b} \\
& & \cr
& &  \mp g r(v^\prime a + a^\prime)  \sqrt {1-a^2} W -a
w^\prime e^{-w}  
-{g}^2 a e^{w} rW^2 +\frac 12 \partial_r [r
e^{-w}(v^\prime a + a^\prime)] =0 \ . \cr
& & \label{c}
\feqn

Similarly from (\ref{tierre}) we have
\eqn
& & \mp \sqrt {1-a^2} Q^s \left[ \frac 12 {g}^2 e^{v+w} W^2 
+ \frac 12 e^{v-w} (v^\prime a + a^\prime)^2 
-\frac 12 \partial_r (v^\prime e^{v-w}) \right] \cr
& &~~~~~~~~~~~~~~~~~~~~~~~~~~
- g^2 a {q^\prime}^X D_X \left( \sqrt {\frac 32} e^v aP^s \right) +\frac {g}2 e^v W^\prime Q^s =0 \ , \label{d}\\
& & \cr
& & \mp g \sqrt {1-a^2} e^v W^\prime +a e^v \partial_r \bigl[e^{-w}
\bigl(v^\prime a + a^\prime\bigr)\bigr] +{g}^2 e^{v+w} W^2 \cr
& &~~~~~~~~~~~~~~~~~~~~~~~~~~~~~~~~~~~ +e^{v-w} \bigl(v^\prime a + a^\prime\bigr)^2 -\partial_r 
\bigl(v^\prime e^{v-w}\bigr)=0 \ . \label{e}
\feqn

In the next section we analyze the system of first order differential equations obtained above. 

\section{\large Static BPS configurations}\label{blackhole}
Let us begin with equation (\ref{a}) from which we easily obtain
\eqn
e^v =\frac {r}{r_0 \sqrt {1-a^2}}  \label{delta}
\feqn
where $r_0$ is an integration constant.
Using (\ref{1fr}) this can be rewritten as
\eqn
1=\mp g r_0 We^{v+w} \label{alfa}
\feqn

We focus on the equations derived in the previous section
to obtain a set of independent ones.
We start with the equation (\ref{b}). Using
(\ref{delta}) we find
\eqn
\partial_r \left( g a e^v W \mp \frac r{2r_0} e^{-w} (v^\prime a + a^\prime) \right) =0
\label{91}
\feqn
It is straightforward to verify that (\ref{91}) is satisfied by (\ref{alfa}) and (\ref{511}).
Thus (\ref{b}) is identically satisfied.

Now we turn to the analysis of the equation (\ref{c}). 
Using
\eqn
rv^\prime a +ra^\prime =a-\frac{ra^\prime}{a^2 -1} \label{lambda}
\feqn
it becomes
\eqn
\partial_r \left( ae^{-w} +\frac 12 r e^{-w} (v^\prime a + a^\prime)\right)=0
\feqn
which is solved by (\ref{511}). 
Thus we conclude that the equation  (\ref{c}) is identically satisfied.
We notice that also the equation (\ref{serve?}) is identically satisfied.
Indeed (\ref{511}) gives
\eqn
1-\frac {r^2}4 e^{-2w} (v^\prime a + a^\prime)^2 = a^2 e^{-2w}
+ae^{-2w} r(v^\prime a + a^\prime)
\feqn
Inserting this expression into (\ref{serve?}) we obtain
\eqn
2{g}^2  r^2  W^2 -e^{-2w} (1+rv^\prime ) +ae^{-2w}
r(v^\prime a + a^\prime) + a^2 e^{-2w}=0
\feqn
Using (\ref{1fr}) in the first term and multiplying by $e^{2w}$
we have
\eqn
1-a^2 - rv^\prime +a r (v^\prime a + a^\prime) =0 \label{eta}
\feqn
which is equivalent to (\ref{delta}).

In a similar way we can study the equation (\ref{e}). 
If we use (\ref{1fr}) in the first term of (\ref{e}) we have
\eqn
& & {g}^2 e^{v+w} rWW^\prime +{g}^2 e^{v+w} W^2 +a e^v \partial_r \bigl(e^{-w} \bigl(v^\prime a 
+ a^\prime\bigr)\bigr)  \cr
& &~~~~~~~~~~~~~~~~~~~~~~~~~~~~~~~~ + e^{v-w} \bigl(v^\prime a + a^\prime\bigr)^2 - \partial_r \bigl(v^\prime e^{v-w}\bigr) =0     \label{102}
\feqn
\newpage
This can be rewritten in the form
\eqn
{g}^2 e^{v+w} rWW^\prime +{g}^2 e^{v+w} W^2 +\partial_r (a e^{v-w} (v^\prime a + a^\prime) -v^\prime 
e^{v-w}) =0
\feqn
If we now multiply by $r$ and then add and subtract the quantity
$ e^{v-w} (a(v^\prime a + a^\prime) - v^\prime)$ and finally use (\ref{eta}) 
we find
\eqn                          
{g}^2 \frac {e^{v+w}}2 \partial_r (r^2 W^2) - e^{v-w} (a(v^\prime a + a^\prime) - v^\prime) +\partial_r 
(e^{v-w} (a^2 -1)) =0
\feqn
Now we use $ g^2 r^2 W^2  = (1-a^2 )e^{-2w}$  and obtain
\eqn
\frac {e^{v+w}}2 \partial_r [(1-a^2 )e^{-2w}] - e^{v-w} (a(v^\prime a + a^\prime) - v^\prime) +\partial_r (e^{v-w} (a^2 -1)) =0
\feqn
This shows that (\ref{e}) is identically satisfied.

At the end we consider the equation (\ref{d}).
Using (\ref{1scomposizione}) we have
\eqn
& & \pm \sqrt {1-a^2} \left[ \frac 12 {g}^2 e^{v+w} W^2 + \frac 12 e^{v-w} 
{\left(v^\prime a + a^\prime\right)}^2 -\frac 12 \partial_r \left(v^\prime e^{v-w}\right) \right]\cr 
& &~~~~~~~~~~~~~~~~~~~~~~~~~~~~~~~+\frac 32 g a \partial_r \left(ae^v W\right) 
+\frac {g}2 e^v W^\prime =0
\feqn
By means of (\ref{e}) this can be written as
\eqn
-3ga\partial_r \left(ae^v W \right) + g a^2 e^v W^\prime  \pm a\sqrt {1- a^2} e^v \partial_r 
\left(e^{-w} \left(v^\prime a + a^\prime \right)\right)=0 \label{i}
\feqn
From (\ref{1fr}) we have $\mp g W =\sqrt {1-a^2} \frac {e^{-w}}r $
so that
\eqn
\pm g W^\prime =\frac {a a^\prime}{\sqrt {1-a^2}} \frac {e^{-w}}r 
+\frac {\sqrt {1-a^2}}{r^2} e^{-w} -\frac {\sqrt {1-a^2}}r
\partial_r e^{-w}
\feqn
and then
\eqn
& & g ae^v W^\prime \pm \sqrt {1-a^2} e^v \partial_r (e^{-w} (v^\prime a + a^\prime))
=\pm \frac {v^\prime a}r \sqrt {1-a^2} e^{v-w} \mp \frac ar \sqrt {1-a^2}
e^v \partial_r e^{-w} \cr
& &~~~~~~~~~~~~~~~~~~~~~~~~~~~~~~~~~~~~~~~~~~~~~~~~~~~~~~ \pm \sqrt {1-a^2} e^v \partial_r \left(e^{-w} \left(v^\prime a + a^\prime\right)\right) \cr
& &~~~~~~~~~~~~ = \pm \frac {\sqrt {1-a^2}}r \left( re^v \partial_r (e^{-w} (v^\prime
a + a^\prime)) +e^v v^\prime ae^{-w} -ae^v \partial_r e^{-w} \right) \cr
& &~~~~~~~~~~~~  = \pm \frac {\sqrt {1-a^2}}r e^v \left( -a \partial_r e^{-w} -e^{-w}
a^\prime + \partial_r (r e^{-w} (v^\prime a + a^\prime)) \right)  \cr
& &~~~~~~~~~~~~  = \mp \frac {\sqrt {1-a^2}}r e^v \partial_r \left( ae^{-w} -r e^{-w}
(v^\prime a + a^\prime) \right) =\mp \frac 3{r_0} \partial_r \left(ae^{-w}\right)
\feqn
where in the last step we have used (\ref{511}) and (\ref{delta}). \\
Then using $\frac{e^{-w}}{r_0} = \mp g_R e^v W$ we see that (\ref{i}) is identically satisfied.

In the next section we verify that BPS solutions we have obtained so far satisfy the equations of motion.

\section{\large Equations of motion}\label{equamotion}

The equations of motion of our system are given by
\eqn
& & -e^{v-w} \partial_r (v^\prime e^{v-w}) -3\frac {v^\prime}r  e^{2(v-w)}
+e^{2(v-w)} (v^\prime a + a^\prime)^2 + 4 g^2 e^{2v} W^2  \cr
& &~~~~~~~~~~~~~~~~~~~~~~~~~~~~~~~~~~~~~~~~~~~~~ -\frac 12 g^2 e^{2v} |K|^2 + \frac 32 g^2 e^{2v} 
|K|^2 a^2 =0 \ , 
\label{1} \\
& & \cr
& & e^{w-v} \partial_r (v^\prime e^{v-w}) -\frac 3r w^\prime +|q^\prime |^2
- (v^\prime a + a^\prime)^2 - \frac 12 e^{2w}(8g^2 W^2 + g^2 |K|^2) =0 \ , \label{2} \\ 
& & \cr
& & re^{-2w} \bigl(v^\prime -w^\prime \bigr) -2\bigl(1-e^{-2w}\bigr) +\frac 12 r^2 e^{-2w} 
\bigl(v^\prime a + a^\prime \bigr)^2 \cr
& & ~~~~~~~~~~~~~~~~~~~~~~~~~~~~~~~~~~~~~~~~~~~~~
+ \frac 23 r^2 \left( -6g^2 W^2 +\frac 34 g^2 |K|^2
\right) =0 \ , \label{3} \\
& & a e^v K_X q^{\prime X} =0 \ , \label{4} \\
& & \cr
& & e^{-(v+w)} r^{-3} \partial_r \left(r^3 e^{v-w} {q^\prime}^Z \right)-\frac 12
\partial^Z g_{XY} e^{-2w} q^{\prime X} q^{\prime Y}  \cr
& &~~~~~~~~~~~~~~~~~~~~~~~~~~~~~~~~~ +\frac 34 g^2 a^2
\partial^Z |K|^2 - \partial^Z \left( -6g^2 W^2 +\frac 34 g^2 |K|^2 \right) = 0
\ , \label{5} \\
& & 3e^{-2w} (v^\prime a + a^\prime) +r\partial_r (e^{-w} (v^\prime a + a^\prime))
= g r ae^w |K|^2  \ , \label{6} \\
& & \cr
& & g e^{-2w} K_X q^{\prime X} =0 \ . \label{7}
\feqn
where (\ref{1}), (\ref{2}), (\ref{3}) and (\ref{4}) are the Einstein equations, (\ref{5}) is
the equation for the
scalar fields and (\ref{6}), (\ref{7}) are the Maxwell equations.

First we observe that both (\ref{4}) and (\ref{7}) are an immediate consequence of (\ref{ok}). 
Then we consider the sum of (\ref{1}) and (\ref{2}). Multiplying by $e^{2(v-w)}$ and using (\ref{5restrizione})
we obtain
\eqn
g^2 |K|^2 = 2\frac {e^{-2w}}r (v^\prime +w^\prime ) \ . \label{8}
\feqn
This is solved by (\ref{alfa}), (\ref{1fr}) and (\ref{corol}).

It is straightforward to check that all the equations are indeed satisfied:\\
Equation (\ref{1}) is solved using (\ref{8}), (\ref{102}),
(\ref{alfa}), (\ref{1fr}), (\ref{511}) and (\ref{delta}). \\
Equation (\ref{3}) is solved by (\ref{8}) and (\ref{serve?}).\\
Equation (\ref{5}) is solved using (\ref{restrizione}), (\ref{8}),
(\ref{7restrizione}), (\ref{1fr}) and (\ref{delta}). \\
Finally (\ref{6}) is solved by (\ref{511}) and (\ref{8}).

\section{\large The Universal Hypermultiplet case}\label{solution}

Now we collect the set of first order differential equations obtained by the BPS conditions:
\begin{gather}
1=a^2 e^{-2w} \left[ 1 +\frac {r}2 
\left( v^\prime + \frac {a^\prime} a \right) \right]^2  \\
e^v =\frac {r}{r_0 \sqrt {1-a^2}} \label{def_v}\\
\sqrt {1-a^2} =\mp g e^w rW \label{def_w}\\
{q^\prime}^Z =\pm 3ge^w \sqrt {1-a^2} \partial^Z W 
\end{gather}
It is easy to reduce the above system to
\eqn
\begin{cases}
{q^\prime}^Z = -3 \frac{1-a^2}r \partial^Z \ln W & \text{ }  \\
1-a^2 = \frac{g^2}4 {\left(3 a + r \frac{a^\prime}{1-a^2}\right)}^2 r^2 W^2
& \text{ }  \label{problem}
\end{cases}
\feqn
(\ref{def_v}) and (\ref{def_w}) simply define 
$v$ and $w$ respectively in terms of $a$ and of the scalars, hence as functions of $r$.

In this form our problem is analogous to the domain-wall case: the main difference
is that now the two differential equations are coupled equations and therefore finding an
explicit solution is more involved.
To solve (\ref{problem}) we have to specify $W$ as a function of the scalars
$q^X$ i.e we have to choose which isometry of the quaternionic manifold represents
the action of the $U(1)$ gauge symmetry.
As we have already argued in the previous sections the most interesting
configurations are obtained for isometries in the isotropy group of some point of
the quaternionic manifold. By definition this choice corresponds to a
fixed point solution i.e one with asymptotic constant scalars.
For (supersymmetric) black hole solutions it implies the existence of 
an horizon. The scale invariance appearing in the
near-horizon limit gives rise to the enhancement of the unbroken
supersymmetry associated to a fixed point.

In fact the configurations that have the most relevant role in the AdS/CFT correspondence
are the ones with two fixed points: this type of solutions should describe a
RG flow between two different CFT's defined on the boundary of the five dimensional
space-time. For black hole solutions this means that the space-time is maximally
symmetric in the $r \rightarrow \infty$ limit (for example AdS).
Since as shown in \cite{Alekseevsky:2001if}, 
in order to obtain such configurations one needs the introduction of vector multiplets, 
we postpone the study of black hole configurations to future work.

Here, as an example, we construct an explicit solution of the BPS equations in the case of a
$n_H =1$ hypermultiplet, i.e. the so called universal hypermultiplet.
This simple example is however important because 
this hypermultiplet (which contains the Calabi-Yau volume) appears in any Calabi-Yau compactification.
We adopt for this manifold the notations and the coordinate system of \cite{ceresole:2001}. 
\newpage 
\noindent
For simplicity we make the following choices for the gauged isometry and the graviphoton: 
\begin{gather}
K\equiv \vec{k}_1 = \left(
\begin{array}{c}
0 \\ 1 \\ 0 \\ 0
\end{array}\right) \label{s1}\\  
a^\prime \equiv 0 \label{s2}
\end{gather}

The Killing vector $\vec{k}_1$ has a simple interpretation: it represents the translation of 
the axionic scalar ($\sigma$ in our notations). This solution has been already considered 
in  \cite{gutperle:2001}\footnote{As discussed in section \ref{model} in  \cite{gutperle:2001} 
a mistake affects the calculations; however the final solution has the right functional 
form.}. We remark that since $\vec{k}_1$ is not 
in the isotropy group of any point of the manifold, the presence of fixed points is excluded. 
In general it is easy to see that fixed point solutions are ruled out 
by the assumption $a^\prime=0$. 

We observe that, being the superpotential 
for $\vec{k}_1$  $W_{\vec{k}_1}=\mp \sqrt{\frac 23} \frac 1{4V}$, the only dynamical scalar 
is $V$. The others are space-time constants and can be set consistently equal to zero.  
Imposing (\ref{s1}) and (\ref{s2}) the system (\ref{problem}) becomes
\eqn
\begin{cases}
V^\prime= 6 \frac{1-a^2}r V \label{p1}\\
\frac{1- a^2}{a^2}= \frac 3{32} {\left( \frac{g r}V \right)}^2 \label{p2}
\end{cases}
\feqn
Since the metric of the quaternionic manifold (\ref{quatmetric}) in our parametrization is singular  for $V=0$ 
we restrict ourselves to the branch $V>0$.  
From (\ref{p1}) it follows that the scalar $V$ has the form
\eqn
V={\cal C}\ r^\Lambda
\feqn
with  ${\cal C}$ and $\Lambda={6(1-a^2)}$ fixed,  by consistency with (\ref{p2}), to be
\begin{gather}
\Lambda=1\\
{\cal C}= \sqrt{\frac{3}{32}} \sqrt{\frac{a^2}{1- a^2}} g 
\end{gather}
that in particular gives $a=\sqrt{\frac 56}$.
At the end we find
\begin{gather}
V=\sqrt{\frac {15}{32}} g r 
\end{gather} 
and using (\ref{def_v}) and (\ref{def_w})
\begin{gather}
e^v=\sqrt 6 \frac r{r_0} \\
e^w=\sqrt{\frac {15}8}
\end{gather}
Rescaling the time coordinate $t$ by the constant $\frac{\sqrt 6}{r_0}$ the space-time metric  and 
the electrostatic potential become
\begin{gather}
ds^2 =- r^2 dt^2 + \frac {15}8 dr^2 + r^2 
(d\theta^2 +\sin^2 \theta d\phi^2 +\cos^2 \theta d\psi^2 ) \\
A_t= \sqrt{\frac 54} r
\end{gather}
We notice that as expected the gauging constant $g$ appears only in the 
quaternionic scalar while the other space-time quantities do not depend on it.

\section{\large Discussion and Outlook} \label{conclusion}

In this work we have studied electrostatic spherical BPS (${\cal N}=1$) solutions in ${\cal N}=2$ 
gauged supergravity in five dimensions with hypermultiplet couplings. 
In particular we have discussed the possibility to find  (extreme) black-hole solutions. The main results we have obtained 
can be summarized as follows:
first of all we have derived the BPS equations in a general setting, 
going beyond special cases treated previously with restrictive  ansatz \cite{gutperle:2001}. 

Then we have discussed the structure of the integrability conditions and in particular 
of the hyperini equations. We have obtained relations 
which appear to be a generalization of those found in \cite{ceresole:2001} for  flat domain walls. 
This result is somewhat surprising since our ansatz is quite different from the one 
in \cite{ceresole:2001}.
In addition we have considered a 
configuration 
with a non-vanishing gauge field whose presence complicates considerably the structure of the equations. 

We have verified that our BPS solutions satisfy the equations of motion.

The above results leave much space for further studies.
First of all it would be interesting to explore if and under which conditions the structure found 
for the hyperini equation is maintained when more general ansatz are consideredand 
vector multiplets are introduced. In these general settings one would like
to  explore  the existence of black hole solutions leading to nontrivial RG flows. 

Finally it would be interesting to consider such explicit solutions 
and extend them to non extreme black holes along the lines of what
has been done for the case of vector multiplets \cite{notextremal}.

These open problems are currently under investigations.

\vspace{1.1cm}
\noindent
{\bf Acknowledgments}

\noindent
We would like to thank A. Van Proeyen for very helpful discussions and suggestions at various 
stages of this work. 
In addition we thank F.Belgiorno, G. Dall'Agata, D.Klemm and L.Martucci for useful discussions.
We have been partially supported by INFN, MURST, and the
European Commission RTN program HPRN-CT-2000-00113 in which we
are associated to the University of Torino.

\newpage
\begin{appendix}
\section{\large Conventions}

In this appendix we present some definitions and properties that we use in our work.
With 
\eqn \label{quatscalars} q^X \, \qquad {\scriptstyle X}=1, \ldots , 4 n_H 
\feqn 
we denote the scalars ofthe hypermultiplets which are the coordinates of a quaternionic manifold.
We introduce the $4n_H$beins as 
\eqn \label{fielbein} f^{iA}_X (q^Y ) \ ,\qquad i=1,2 \in SU(2) \ 
, A=1,\ldots,2n_H \in Sp(2n_H) \feqn

The splitting of the flat indices
in $i$ and $A$ reflects the factorization of the holonomy group in $USp(2)(\simeq SU(2))\otimes USp(2n_H)$ which is the main feature of those spaces. 
The indices as a consequence of  the symplectic structure are highered and lowered with the antisymmetric 
matrices \eqn
&& \epsilon_{ij} \ , \qquad C_{AB} \\
&& \epsilon_{ij}=\epsilon^{ij} \ , \qquad \epsilon_{12} =1 \\
&& C_{AB} C^{CB} =\delta_A^{\quad C} \ , \qquad C^{AB} =(C_{AB})^* \ .
\feqn
following the NW-SE convention \cite{ceresole:2001}. 

The important relation 
\eqn 
f_{XiC} f_{Yj}^{\ \ C} =\frac 12 \epsilon_{ij} g_{XY} + R_{XYij} 
\label{fquadrato} 
\feqn
can be viewed as  a definition for the quaternionic metric $g_{XY}$ and for the $SU(2)$ curvature $R_{XYij}$.

We use the symbols $p_{Xi}^{\ \ \ j}$ for the $SU(2)$ spin connection
whereas $\omega_\mu^{ab}$ denotes the usual Lorentz spin connection.
The covariant derivative which appears in the gravitini supersymmetry variation  acts on the  symplectic Maiorana spinors $\epsilon_i$ as
\eqn
\DD_\mu \epsilon_i  &=& \partial_\mu \epsilon_i +\frac 14
\omega^{ab}_\mu \gamma_{ab} \epsilon_i -\partial_\mu q^X p_{Xi}^{\ \ j}
\epsilon_j 
- g A_\mu P_i^{\ j} \epsilon_j
\feqn
where  the generalized spin connection receives the following contributions: the first term represents 
the Lorentz
action while the others can be identified  with the $SU(2)$ action plus a term due  
to the R-symmetry $SU(2)$ gauging. $A_{\mu}$ is the graviphoton $1-$form and $P^r$
are the prepotentials while $g$ is the electric gauge coupling.

It is useful to introduce the projection on the Pauli matrices for quantities in the adjoint representation of $SU(2)$, for example
\eqn
R_{XYi}^{\qquad j}=R_{XY}^r (i\sigma_r)_i^{\ j} \label{pro}
\feqn
where $(\sigma_r)_i^{\ j}$ are the usual Pauli matrices
\eqn
\sigma_1 = \left( \begin{array}{cc} 0 & 1 \\ 1 & 0 \end{array} \right) \ ,
\qquad
\sigma_2 = \left( \begin{array}{cc} 0 & -i \\ i & 0 \end{array} \right) \ ,
\qquad
\sigma_3 = \left( \begin{array}{cc} 1 & 0 \\ 0 & -1 \end{array} \right) \ .
\feqn
which satisfy
\eqn
& & (\sigma_r)_i^{\ j} (\sigma_s)_j^{\ k} =\delta_{rs} \delta_i^{\ k}
+i\epsilon_{rs}^{\quad t} (\sigma_t)_i^{\ k} \label{prodotto} \\
& & \left[\sigma_r ,\sigma_s \right] =2i
\epsilon_{rst} \sigma^t \label{parentesi}
\feqn
The prepotentials are defined by the relation
\eqn
& & R_{XY}^r K^Y = D_X P^r \label{prepot} \\
& & D_X P^r := \partial_X P^r +2\epsilon^{rst} p_X^s P^t 
\feqn 
where $D_X$ is the $SU(2)$ covariant derivative.
They can be expressed in terms of the Killing 
vectors 
\eqn P^r =\frac 1{2n_H} D_X K_Y R^{rXY} \label{killpot} 
\feqn 

The scalar potential is defined as 
\begin{equation}
{\cal V}= -4 P^r P^r g^2 +2 N_{Ai}N^{Ai} g^2 \label{scalarpot}\end{equation} with
\begin{equation}
 N^{Ai}= \frac {\sqrt 6}4 K^X 
f^{Ai}_X = \frac 2{\sqrt 6} f^{Ai}_X R^{rYX} D_Y P^r \end{equation}
Defining the superpotential $W$ by $P^r= \sqrt{\frac 32} W Q^r$ the potential becomes 
\eqn
{\cal V} = -6 g^2 W^2  + g^2 \frac92 g^{XY} \partial_X W \partial_Y W
\feqn
The universal hypermultiplet ($n_H=1$) corresponds to the quaternionic K\"ahler space $\frac{SU(2,1)}{SU(2)\times
U(1)}$. A significant parametrization, from a M-theory point of view, is \cite{ceresole:2001}
$$q^X= \{V,\sigma, \theta, \tau\}$$ with the metric
\begin{equation}
\rmd  s^2 = \frac{\rmd V^2}{2V^2} + \frac{1}{2V^2}\left( \rmd \sigma + 2
\theta \, \rmd  \tau - 2 \tau \, \rmd  \theta\right)^2 + \frac{2}{V} \,
\left(\rmd \tau^2 + \rmd  \theta^2\right) \,. \label{quatmetric}
\end{equation}
Using the general properties of quaternionic geometry it is possible from (\ref{quatmetric}) to derive explicitly all the quantities presented above, in particular the Killing vectors and the prepotentials of the eight isometries of manifold. For the axionic shift we have:
\begin{equation}
\vec{k}_1 = \left(
\begin{array}{c}
0 \\ 1 \\ 0 \\ 0
\end{array}
\right)\,
~~~~~~~~~~~~~~~~~~~~~~~~~~~~~~~~~~~~~
\vec{P}_1 = \left(
\begin{array}{c}  0 \\ 0  \\ -\frac{1}{4V} \end{array}
\right)\,
\end{equation}

\end{appendix}

\end{document}